\documentclass[preprint,a4paper,superscriptaddress,floatfix,citeautoscript,floatfix]{revtex4}
\usepackage{amsmath,amssymb}
\usepackage{graphicx}

\newcommand{\Id}[1] {\int \! \! {\rm d}^3 #1}

\renewcommand{\vr}{{\bf r}}
\newcommand{\vj}{{\bf j}}

\begin{document}

\title{Toward an all-round semi-local potential for the electronic exchange}

\author{Micael J.\,T. Oliveira}
\email[Electronic address:\;]{micael@teor.fis.uc.pt}
\altaffiliation[Present address: ]{Center for Computational Physics, University of Coimbra, Rua Larga, 3004-516 Coimbra, Portugal}
\affiliation{Laboratoire de Physique de la Mati\`{e}re
Condens\'{e} et Nanostructures,
Universit\'{e} Lyon I, CNRS, UMR 5586, Domaine scientifique de la
Doua, F-69622 Villeurbanne Cedex, France}
\affiliation{European Theoretical Spectroscopy Facility (ETSF)}

\author{Esa R{\"a}s{\"a}nen}
\email[Electronic address:\;]{erasanen@jyu.fi}
\affiliation{Nanoscience Center, Department of Physics, University of
  Jyv\"askyl\"a, FI-40014 Jyv\"askyl\"a, Finland}

\author{Stefano Pittalis}
\affiliation{Department of Physics and Astronomy, University of Missouri, Columbia, Missouri 65211, USA}

\author{Miguel A.\,L. Marques}
\affiliation{Laboratoire de Physique de la Mati\`{e}re
Condens\'{e} et Nanostructures,
Universit\'{e} Lyon I, CNRS, UMR 5586, Domaine scientifique de la
Doua, F-69622 Villeurbanne Cedex, France}
\affiliation{European Theoretical Spectroscopy Facility (ETSF)}

\begin{abstract}
  We test local and semi-local density functionals for the electronic
  exchange for a variety of systems including atoms, molecules, and
  atomic chains. In particular, we focus on a recent universal
  extension of the Becke-Johnson exchange potential [R\"as\"anen,~E.;
  Pittalis,~S.; Proetto,~C. R. \emph{J. Chem. Phys.}  \textbf{2010},
  \emph{132}, 044112]. It is shown that when this potential is used
  together with the Becke-Roussel approximation to the Slater
  potential [Becke,~A.~D.; Roussel,~M.~R. \emph{Phys. Rev. A}
  \textbf{1989}, \emph{39}, 3761--3767], a good overall agreement is
  obtained with experimental and numerically exact results for several
  systems, and with a moderate computational cost. Thus, this
  approximation is a very promising candidate in the quest for a
  simple and all-round semi-local potential.
\end{abstract}


\maketitle

\section{Introduction}

Density-functional theory~\cite{dg90,barth04:9} (DFT) has become the
standard tool both in quantum chemistry and in atomic, molecular, and
solid-state physics.  The practical applicability of DFT crucially
depends on the approximation for the exchange-correlation (xc) energy
functional. The ``Jacob's ladder'' of functionals developed in the
past few decades~\cite{pk03} has posed the following well-known
problem: by climbing successive rungs of the ladder one increases the
accuracy of the functional, but one also increases substantially the
computational burden of the method.  Finding a balance between
accuracy and efficiency, together with {\em universality} (which is
the ideal ability to deal equally well with any kind of system), has
remained a major challenge in DFT.

As the simplest density functionals, occupying the first two rungs of
Jacob's ladder, the local density approximation (LDA) and
generalized-gradient approximations (GGA) are numerically efficient
and surprisingly accurate for many (strongly inhomogeneous)
systems. However, both these families of functionals exhibit well-known
failures in the calculation of, e.g., band gaps of semiconductors and
insulators~\cite{hpsm05:174101}, the response to electric
fields~\cite{cpgbssrk98:10489}, etc.  The problems are particularly
dramatic in systems where long-range interactions play a 
crucial role, i.e., elongated molecules and atomic
chains~\cite{kkp04:213002,gggb02:9591,mwy03:11001,nf07:191106,kak09:712,rpcsv08:060502,mwdrs09:300,ggb02:6435}.
The main origin for these errors is the wrong (exponential) asymptotic
behavior and the lack of the derivative discontinuity in the xc
potential.

Climbing further the ladder, the optimized-effective-potential (OEP)
method~\cite{sh53:317,ts76:36,kk08:3} or its simplification within the
Krieger-Li-Iafrate (KLI) approximation~\cite{kli92:101} provide, in
principle, an access to the {\em exact} exchange energy and potential
within DFT.  Thus, as long as the electronic correlation is not
significant, the OEP and KLI are free from the failures mentioned
above.  However, as non-local orbital functionals they are
computationally demanding and therefore usable only for systems
containing a small number of particles.

To bridge the gap between the GGA and OEP,
meta-GGAs~\cite{pkzb99:2544,tpss03:146401} are appealing
candidates. They supplement the GGA by further semi-local information 
through the kinetic-energy density and/or the Laplacian of the density, 
and, in some cases, also through the paramagnetic current density. 
Recently, R\"as\"anen, Pittalis, and Proetto~\cite{rpp10:044112} 
(RPP) developed a meta-GGA for the exchange part of the xc potential.
The RPP potential introduces a number of important constraints 
and features (see below), and performs well for, e.g., non-Coulombic 
systems and atomic chains. It is based on the Becke-Johnson 
(BJ) potential~\cite{bj06:221101} -- a simple meta-GGA 
close to the OEP accuracy for atoms -- but,  in contrast to BJ, 
the RPP potential is fully
gauge-invariant, exact for any one-particle system, and has the
correct asymptotic behavior for any $N$-particle system. 

Also other modifications to the BJ potential have been
suggested to improve the performance for atomic
chains~\cite{akk08:165106} and band gaps~\cite{tb09:226401}. In fact,
the latter modification~\cite{tb09:226401} allows the calculation of
band gaps of semiconductors and insulators with an error of the same
order of GW calculations, but at a very small fraction of the GW
computational time.

In this paper we test the RPP potential~\cite{rpp10:044112}, used
together with the Becke-Roussel (BR) approximation to the Slater
potential~\cite{br89:3761}, for a large variety of systems. 
We compare this approximation
to the BJ one, also complemented by the BR potential.
In order to allow for a comparison to experimental reference data, 
we have added to the above exchange potentials the correlation within 
the LDA. We compare the results also against the LB94 potential of 
van Leeuwen and Baerends~\cite{lb94:2421} (a GGA with correct asymptotic 
behavior including also correlation). Moreover, for completeness, we
include results calculated with standard LDA and GGA functionals. As a
reference we use experimental or high-quality ab initio data. In some
cases the performance of the exchange potentials alone, i.e., 
without the addition of correlation, is compared to the 
exact-exchange OEP results. The combination of RPP and BR potentials
is found to yield the best overall performance of the tested 
approximations, and thus it provides a promising step toward an
all-round semi-local exchange potential in DFT.

\section{Theory}

\subsection{Exact exchange}
\label{subsec:exx}

In a majority of atomic, molecular, and solid-state systems, the electronic
exchange gives, in absolute terms, a much larger contribution to (most) observables 
than the correlation. Therefore, in practical applications the exchange 
is the most important term to be approximated in the functional.
The exact exchange energy in Hartree atomic units (a.u.) is written as
\begin{equation}
  E_\text{x}[\rho_{\sigma}] = -\frac{1}{2} \sum_{\sigma=\uparrow,\downarrow} \sum_{j,k=1}^{N_\sigma} 
  \Id{r}\!\!\Id{r'} 
  \frac{\varphi_{j \sigma}^*(\vr)\varphi_{k \sigma}^*(\vr')
  \varphi_{j \sigma}(\vr')\varphi_{k \sigma}(\vr)}{|\vr
  - \vr'|} \; ,
  \label{Ex}
\end{equation}
and its functional derivative gives the Kohn-Sham (KS) exchange
potential as $v_{\text{x}\sigma}(\vr)=\delta
E_\text{x}/\delta\rho_\sigma(\vr)$. These quantities can be rigorously
calculated with the OEP method~\cite{sh53:317,ts76:36,kk08:3} through
an integral equation that has to be solved together with the KS
equations.

At this point, it is useful to write the (KS) exchange potential as a sum
\begin{align}
  v_{\text{x} \sigma}(\vr) = & v^\text{SL}_{\text{x} \sigma}(\vr) + \Delta v^\text{OEP}_{\text{x} \sigma} (\vr) \;\nonumber \\
  = &
  v^\text{SL}_{\text{x} \sigma}(\vr) + \Delta v^\text{KLI}_{\text{x} \sigma}(\vr) + \Delta v^\text{OS}_{\text{x} \sigma}(\vr)\;,
  \label{vx}
\end{align}
where 
\begin{equation}
  v^{\rm SL}_{\text{x}\sigma}(\vr) = -\sum_{j,k=1}^{N_\sigma} 
  \Id{r'} \frac{\varphi_{j \sigma}^*(\vr)\varphi_{k \sigma}^*(\vr')
  \varphi_{j \sigma}(\vr')\varphi_{k \sigma}(\vr)}{\rho_\sigma(\vr) |\vr
  - \vr'|} \; ,
  \label{slater}
\end{equation}
is the Slater potential, i.e., the average of the Fock potential felt
by the electrons, and $\Delta v^\text{OEP}_{\text{x} \sigma} (\vr)$ is
the exact (OEP) contribution~\cite{sh53:317,ts76:36,kk08:3}, which can
be decomposed into the Krieger-Li-Iafrate~\cite{kli92:101} (KLI) part
and the orbital shifts.  Apart from, e.g., atomic
chains~\cite{kkp04:213002}, the orbital shifts in a ground-state
calculation are usually of minor importance and therefore neglected,
leading to so-called KLI approximation. This relieves the
computational burden of solving the integral equation, but the tedious
integrals in the Slater potential are still to be
calculated. Therefore, even within the KLI approximation the
efficiency of an OEP calculation is far from that of semi-local
functionals.

\subsection{Becke-Johnson potential}

The BJ potential~\cite{bj06:221101} is a simple approximation to the
OEP contribution in Eq.~(\ref{vx}),
\begin{equation}
  \Delta v^{\text{OEP}}_{\text{x} \sigma} (\vr)\approx \Delta v^{\text{BJ}}_{\text{x} \sigma} (\vr) = 
  C_{\Delta v} \sqrt{ \frac{ \tau_{\sigma}(\vr) }{ \rho_\sigma(\vr) }},
\end{equation}
where
\begin{equation}\label{stau}
\tau_\sigma(\vr)=\sum_{j=1}^{N_\sigma} |\nabla\varphi_{j\sigma}(\vr)|^2
\end{equation}
is (twice) the spin-dependent kinetic-energy density, and $C_{\Delta
  v} = \sqrt{5/(12\pi^2)}$. The BJ potential is exact for the hydrogen
atom and for the homogeneous electron gas, and, regarding quantum
chemistry applications, it has several beneficial properties.  First,
it yields the atomic step structure in the exchange potential (which
was the main motivation for the approximation) very
accurately~\cite{bj06:221101}. Secondly, it has the derivative
discontinuity for fractional particle numbers~\cite{akk08:165106}.

To improve on the numerical efficiency of this potential, one often
replaces also the Slater potential $v^{\rm SL}_{\text{x}\sigma}(\vr)$
by the Becke-Roussel potential~\cite{br89:3761}. This is again a
meta-GGA potential, written in terms of $\nabla^2 \rho_\sigma$ and of
$\tau_\sigma$, that reproduces to a very high precision the Slater
potential for atoms.

\subsection{Universal extension to Becke-Johnson}

The main limitations of the BJ potential are the facts that it is not
gauge-invariant and that it is not exact for {\em all} one-particle
systems. Both limitations were recently removed in the extension by
RPP~\cite{rpp10:044112}, that proposed the form
\begin{equation}
  \Delta v^\text{OEP}_{\text{x} \sigma} (\vr)\approx \Delta v^\text{RPP}_{\text{x} \sigma} (\vr) = 
  C_{\Delta v} \sqrt{\frac{ D_{\sigma}(\vr) }{ \rho_\sigma(\vr) }},
\end{equation}
where
\begin{equation}
  \label{D}
  D_{\sigma}(\vr)= \tau_{\sigma}(\vr)-\frac{1}{4}\frac{\left[ \nabla \rho_\sigma(\vr)
  \right]^2}{\rho_\sigma(\vr)}-\frac{\vj^2_{p \sigma}(\vr)}{\rho_\sigma (\vr)},
\end{equation}
describes the local curvature of the exchange (Fermi)
hole~\cite{dobson93:8870}. This quantity has already been useful in
the derivation of several
functionals~\cite{br89:3761,becke88:1053,becke96:995,prhg07:235314,prpg09:085316,rpp10:195103}
and is the key ingredient of the electron-localization
function~\cite{bh90:5397,bmg05:010501,rcg08:115108}, a standard tool
used to analyze bonding in electronic systems.  Finally, the
spin-dependent paramagnetic current density is defined as
\begin{equation}
  \label{current}
  \vj_{p \sigma}(\vr)=\frac{1}{2i}\sum_{j=1}^{N_\sigma} \left\{
   \varphi^*_{j \sigma}(\vr) \left[\nabla \varphi_{j \sigma}(\vr)\right] - \left[\nabla \varphi^*_{j \sigma}(\vr)\right] 
  \varphi_{j \sigma}(\vr) \right\}.
\end{equation}
The RPP approximation is gauge-invariant and it is exact for {\em all}
one-particle systems. Furthermore, it has a correct asymptotic limit
for finite $N$-electron systems (except on nodal surfaces of the
energetically highest-occupied
orbitals~\cite{sg02:033003,kp03:035103}). The universality of the
approach, whose principles has also been shown to work in two
dimensions~\cite{prp10:115108}, is reflected into a resulting
potential that can be applied reasonably well to any kind of system.
For example, the RPP potential has been seen to reproduce well the KLI
potential in hydrogen chains in electric fields and in Hooke's atoms
subject to magnetic fields~\cite{rpp10:044112}. In this respect,
another recent extension of the BJ potential can be viewed as more
restrictive~\cite{kak09:712}. The present study aims at further
evaluating the capability of this approximation for atoms, small
molecules, and atomic chains.

\section{Numerical procedure}

The evaluation of the Slater part in the BJ~\cite{bj06:221101} and
RPP~\cite{rpp10:044112} potentials is computationally more demanding
than the evaluation of the correction terms $\Delta
v^{\text{BJ}}_{\text{x} \sigma}$ and $\Delta v^{\text{RPP}}_{\text{x}
  \sigma}$. Nevertheless, as already pointed out by Becke and
Johnson~\cite{bj06:221101}, it is possible to approximate the Slater
part by using the semi-local Becke-Roussel (BR) exchange-energy
functional~\cite{br89:3761}. In this way, the cost of evaluating the
full BJ and RPP potentials becomes similar to the one of a usual LDA
or GGA. To avoid any ambiguity, we will hereafter denote the BJ and
RPP potentials, where the Slater part was replaced by the BR
potential, as BJBR and RPPBR, respectively

When using experimental results as a reference, it is necessary to add a
correlation contribution to the BJBR and RPPBR potentials for a proper
comparison. We use the correlation in the LDA level within
the Perdew-Wang~\cite{pw92:13244} (PW) form. The results are
compared also to the standard LDA -- with the PW parametrization
for the correlation part, the GGA of Perdew, Burke, and
Ernzerhof~\cite{pbe96:3865} (PBE), and the GGA of Leeuwen and
Baerends~\cite{lb94:2421} (LB94) -- again using the PW parametrization
for the LDA part of the potential. In all cases we have applied the
potentials self-consistently in the KS-DFT framework. Although in the
case of PBE the correlation functional used is not the same as in the
other cases, we expect this fact to result in negligible differences
in the quantities and systems studied in this work.

In the case of atoms and hydrogen chains, calculations are also
performed using {\em exchange-only} potentials. Results are then
compared with exact-exchange OEP data available in the literature. 
Besides the BJBR and RPPBR potentials, we also performed these 
calculations using the exchange part of the LDA (xLDA) and of 
the PBE (xPBE).

It is important to bear in mind that BJ, RPP, and LB94 are such
approximations to the exchange (or xc) potential that are not
functional derivatives of corresponding exchange (or xc) {\em
  energies}~\cite{gs09:044107}.  Here we focus on fairly standard
quantities that may be accessed without the computation of total
energies. These quantities includes ionization potentials and
electronic affinities of atoms, ionization potentials and dipole
polarizabilities of small molecules, and longitudinal polarizabilities
of hydrogen chains. We believe that these benchmarks provide us with a
fairly complete view on the properties of different approximations
considered in this work.

All the single-atom calculations are performed with the {\sc APE}
code~\cite{on08:524}, while molecules and atomic chains are dealt with
{\sc octopus}~\cite{mcbr03:60,caoralmgr06:2465}. In the latter case,
the electron-ion interaction is handled through norm-conserving
pseudopotentials generated with {\sc APE} for each functional and
approximation studied in this work.

\section{Results}

\subsection{Atoms}

First we consider single atoms and focus on the ionization energies
and electron affinities (see \ref{table_atoms1}).  There are several
ways to estimate these quantities within DFT. The most direct one is
to calculate the differences in total energy of both the neutral atom,
and of its anion and cation, respectively. In this way, traditional
LDA and GGA functionals usually yield quite good ionization
potentials. Electron affinities are more complicated as often LDAs and
GGAs fail to bind the extra electron.

The other approach, the one used in this work, is to look at the KS
eigenenergy of the highest occupied atomic orbital (HOMO), that should
be equal to the negative of the ionization potential. The electron
affinity is computed simply from the ionization potential of the
respective anion. This method samples much better the quality of the
potential, and it is particularly sensitive to the asymptotic
description of the potential.

As known from previous studies~\cite{lb94:2421}, the LDA and PBE
perform poorly for the ionization potential: the mean absolute error
(last row of \ref{table_atoms1}) is larger than 40\% for this set of
atoms. The result indicates the crucial role of the correct asymptotic
behavior in the exchange potential. The decay of the xc potential is
properly described by LB94 potential showing good performance.  For
the same reason, good results have been obtained also with KLI-CS -- a
combination of KLI~\cite{kli92:101} for the exchange and the
Colle-Salvetti~\cite{cs75:329} functional for the correlation -- as
reported by Grabo and Gross~\cite{gg95:141}.  It seems that RPPBR is
slightly more accurate than the original BJBR potential, also when
considering the exchange potentials alone.  When compared against
exact-exchange OEP results~\cite{ev93:2800}, xLDA and xPBE perform
poorly, while BJBR and RPPBR perform better, the later being now more
accurate.

\begin{turnpage}
\begin{table}[ht]
  \caption{Ionization potentials from the highest occupied Kohn-Sham orbital (in a.u.)$^a$
    \label{table_atoms1}}

  \begin{tabular}{l c c c c c c c c c c c c}
    \hline
  Atom&  xLDA&  xPBE&  BJBR& RPPBR& OEP$^b$&   LDA&   PBE&  LB94&KLI-CS$^c$& BJBR-PW & RPPBR-PW & Expt.$^d$ \\
    \hline
    He& 0.517& 0.553& 0.857& 0.924& 0.918& 0.570& 0.585& 0.851&	0.945&	0.922& 0.982& 0.903 \\
    Li& 0.100& 0.109& 0.254& 0.183& 0.196& 0.116& 0.111& 0.193&	0.200&	0.276& 0.201& 0.198 \\
    Be& 0.170& 0.182& 0.355& 0.300& 0.309& 0.206& 0.201& 0.321&	0.329&	0.401& 0.338& 0.343 \\
    B & 0.120& 0.128& 0.279& 0.226&      & 0.151& 0.143& 0.296&	0.328&	0.321& 0.260& 0.305 \\
    C & 0.196& 0.204& 0.399& 0.332&      & 0.227& 0.218& 0.401&	0.448&	0.440& 0.366& 0.414 \\
    N & 0.276& 0.285& 0.526& 0.451& 0.571& 0.309& 0.297& 0.510&	0.579&	0.567& 0.486& 0.534 \\
    O & 0.210& 0.224& 0.391& 0.383&      & 0.272& 0.266& 0.516&	0.559&	0.472& 0.450& 0.500 \\
    F & 0.326& 0.339& 0.564& 0.526&      & 0.384& 0.376& 0.647&	0.714&	0.636& 0.588& 0.640 \\
    Ne& 0.443& 0.456& 0.743& 0.686& 0.851& 0.498& 0.491& 0.788&	0.884&	0.810& 0.745& 0.792 \\
    Na& 0.097& 0.103& 0.247& 0.178& 0.182& 0.113& 0.106& 0.205&	0.189&	0.270& 0.197& 0.189 \\
    Mg& 0.142& 0.149& 0.313& 0.252& 0.253& 0.175& 0.168& 0.291&	0.273&	0.357& 0.287& 0.281 \\
    Al& 0.086& 0.092& 0.227& 0.160&      & 0.111& 0.102& 0.216&	0.222&	0.263& 0.188& 0.220 \\
    Si& 0.144& 0.150& 0.320& 0.237&      & 0.170& 0.160& 0.290&	0.306&	0.356& 0.267& 0.300 \\
    P & 0.203& 0.210& 0.416& 0.324& 0.392& 0.231& 0.219& 0.369&	0.399&	0.453& 0.355& 0.385 \\
    S & 0.174& 0.182& 0.349& 0.305&      & 0.229& 0.219& 0.410&	0.404&	0.420& 0.362& 0.381 \\
    Cl& 0.254& 0.262& 0.469& 0.400&      & 0.305& 0.295& 0.491&	0.506&	0.533& 0.453& 0.477 \\
    Ar& 0.334& 0.343& 0.592& 0.506& 0.591& 0.382& 0.373& 0.577&	0.619&	0.652& 0.557& 0.579 \\
    \hline
    $\Delta\,(\%)$& 43& 41& 13.8& 8.5& & 41& 42& 3.7& 5.7& 14.4& 7.4& \\
    \hline
  \end{tabular}
  \begin{flushleft}
    $^a$ The last row shows the mean absolute error in percentage with
    respect to exact-exchange and experimental results for exchange
    potentials and combined exchange and correlation potential,
    respectively. $^b$ From the work of Engel and
    Volko.~\cite{ev93:2800} $^c$ From the work of Grabo and
    Gross.~\cite{gg95:141} $^d$ Experimental results taken from Ratzig
    and Smirnov~\cite{rs85}.
  \end{flushleft} 
\end{table}
\end{turnpage}

As noted already by Becke and Johnson~\cite{bj06:221101}, the BJ
exchange potential goes asymptotically to a finite (non-zero)
constant.  In principle, this constant only redefines the zero of
orbital energy, and should have no implication in the quality of the
results, but it has to be taken into account when computing the
ionization potential. This can be done by subtracting the value of the
constant, which can be obtained from the asymptotic expansion of the
density and the kinetic energy density, to the value of the KS
eigenenergy of the HOMO. A perfectly equivalent procedure is to shift
the BJ exchange potential so that it goes asymptotically to zero. In
the case spin-uncompensated atoms the constant depends on spin. Then
it is possible to shift the spin-up and spin-down potentials by
different amounts, provided that this does not imply a change in the
occupancies of the orbitals.

\begin{table}
  \caption{\label{table_atoms2} Electron affinities calculated from
    the highest occupied Kohn-Sham orbital of the anion (in a.u.)$^a$
  }
\begin{tabular}{l c c c c c}
\hline
Atom    &LB94   &KLI-CS$^b$ &BJBR-PW &RPPBR-PW &Expt.$^c$  \\
\hline
Li	&0.020	&0.024	&-      &0.036	&0.023 \\
B	&0.016	&0.033	&-      &-	&0.010 \\
C	&0.049	&0.083	&-      &0.032	&0.046 \\
O	&0.077	&0.110	&-      &-	&0.054 \\
F	&0.128	&0.208	&-      &0.110	&0.125 \\
Na	&0.023	&0.022	&0.012  &0.036	&0.020 \\
Al	&0.018	&0.024	&-      &-	&0.016 \\
Si	&0.050	&0.065	&0.019  &0.039	&0.051 \\
P	&0.061	&0.048	&-      &0.026	&0.027 \\
S	&0.098	&0.106	&-      &0.069	&0.076 \\
Cl	&0.140	&0.174	&0.118  &0.127	&0.133 \\
\hline
$\Delta\,(\%)$  & 29  & 66 & 38$^d$ &  28$^d$ & \\
\hline
\end{tabular}
  \begin{flushleft}
    $^a$ The last row shows the mean absolute error in percentage.
    $^b$ From the work of Grabo and Gross.~\cite{gg95:141} $^c$
    Experimental results taken from Ratzig and Smirnov~\cite{rs85}.
    $^d$ Mean error calculated for bound solutions only.
  \end{flushleft} 
\end{table}

The electron affinities for our set of atoms are given in
\ref{table_atoms2}. As it is well known, the LDA or most GGAs do not
give bound solutions for most negative ions, so we chose not to
include them in the table.  In most cases BJBR failed to give bound
solutions for the anions, while for RPPBR this happened only in a few
cases. Considering only the cases were RPPBR gave bound solutions, the
deviation from the exact values was around 28\%. It seems that LB94,
having a similar overall accuracy, works better for small ions,
whereas RPPBR increases its accuracy for larger systems. For example,
for the last three atoms in \ref{table_atoms2} (P, S, Cl) RPPBR has an
error of only a few percent. Interestingly, KLI-CS results deviate by
more than 60\% from the exact values. This might be due to the poor
compatibility between the exact non-local exchange and the correlation
part, when the asymptotic regime is strongly dominated by the ionic
HOMO.

\subsection{Molecules}

\begin{table}
  \caption{\label{table_molecules1}
    Ionization potentials for molecules calculated from the
    highest occupied Kohn-Sham orbital (in eV)$^a$
  }
\begin{tabular}{l c c c c c c}
\hline
Molecule & LDA & PBE & LB94 &BJBR-PW &RPPBR-PW & Expt.$^b$\\
\hline
CS$_2$&	6.93	&6.81	&11.54	&13.08	&10.76&	10.07\\
H$_2$S&	6.4	&6.3	&11.33	&12.51	&11.05&	10.46\\
C$_2$H$_4$&	6.92	&6.74	&11.85	&12.71	&10.96&	10.51\\
PH$_3$&	6.69	&6.64	&11.65	&12.88	&11.62&	10.59\\
NH$_3$&	6.28	&6.19	&11.55	&12.58	&11.3&	10.8\\
Cl$_2$&	7.47	&7.36	&12.3	&14.03	&11.86&	11.48\\
C$_2$H$_6$&	8.13	&8.15	&12.94	&15.04&	13.33&	12\\
SiH$_4$&	8.53	&8.53	&13.44	&15.44	&14.04&	12.3\\
SO$_2$&	8.3	&8.09	&14.06	&15.2&	13.29&	12.35\\
H$_2$O&	7.38	&7.23	&13.2	&14.08&	12.66&	12.62\\
HCl& 	8.14	&8.04	&13.29	&14.81&	12.83&	12.74\\
N$_2$O&	8.6	&8.35	&14.48	&15.4&	13.37&	12.89\\
CH$_4$&	9.46	&9.45	&14.29	&16.69&	14.65&	13.6\\
CO$_2$&	9.31	&9.05	&15.32	&16.37&	14.2&	13.78\\
CO&	9.16	&9.09	&14.49	&16.46&	14.47&	14.01\\
H$_2$&	10.28	&10.4	&15.27	&17.92&	17.54&	15.43\\
N$_2$&	10.39	&10.24	&16.94	&18.18&	16.09&	15.58\\
F$_2$&	9.79	&9.54	&17.03	&17.56&	16.18&	15.7\\
HF&	9.85	&9.65	&16.44	&17.3&	15.69&	16.03\\
\hline
$\Delta\,(\%)$  & 35  & 36   &  8.0  & 19    & 5.7   &    \\
\hline
\end{tabular}
  \begin{flushleft}
    $^a$ The last row shows the mean absolute error in percentage.
    $^b$ Experimental results taken from Gr{\"u}ning et
    al.~\cite{gggb02:9591}.
  \end{flushleft} 
\end{table}

Next we test the approximations for a large set of small molecules by
computing ionization potentials and static (isotropic) dipole
polarizabilities. The ionization potentials are obtained from the HOMO
as in the previous section, while the polarizabilites are computed as
a derivative of the dipole moment of the system with respect to the
applied electric field. The ionization potentials are listed in
\ref{table_molecules1}. Interestingly, RPPBR is significantly more
accurate than BJBR and deviates less than 6\% from the experimental
values. LB94 performs also well with a mean absolute error of 8\%. In
contrast, the LDA and PBE fail in a similar fashion as in the atomic
cases considered in the previous section.

\begin{table}
  \caption{\label{table_molecules2}
    Static (isotropic) dipole polarizabilities for molecules (in a.u.)$^a$
  }
\begin{tabular}{l c c c c c c}
\hline
\hline
Molecule & LDA	& PBE & LB94& BJBR-PW & RPPBR-PW & Expt.$^b$\\
\hline
CS$_2$&	56.50&	56.45&	51.72&	55.44&	55.29&	55.28\\
H$_2$S&	26.21&	25.91&	21.95&	24.24&	22.51&	24.71\\
C$_2$H$_4$&	28.71&	28.52&	24.93&	27.71&	25.21&	27.7\\
PH$_3$&	32.29&	31.72&	27.39&	29.99&	27.47&	30.93\\
NH$_3$&	15.58&	15.45&	12.41&	13.83&	12.32&	14.56\\
Cl$_2$&	32.33&	32.21&	30.92&	31.41&	32.39&	30.35\\
C$_2$H$_6$&	30.17&	29.73&	27.41&	28.23&	26.52&	29.61\\
SiH$_4$&	34.03&	33.07&	30.17&	31.11&	28.47&	31.9\\
SO$_2$&	27.44&	27.53&	22.97&	25.78&	23.68&	25.61\\
H$_2$O&	10.74&	10.73&	8.28&	9.49&	8.53&	9.64\\
HCl &	18.61&	18.47&	15.85&	17.18&	16.21&	17.39\\
N$_2$O&	20.7&	20.74&	17.42&	19.46&	18.47&	19.7\\
CH$_4$&	17.77&	17.45&	15.87&	16.46&	15.41&	17.27\\
CO$_2$&	18.21&	18.24&	15.66&	17.39&	16.16&	17.51\\
CO&	13.91&	13.87&	11.6&	13.13&	12.29&	13.08\\
H$_2$&	5.87&	5.64&	5.02&	5.27&	4.56&	5.43\\
N$_2$&	12.64&	12.63&	10.79&	11.9&	11.4&	11.74\\
F$_2$&	8.86&	8.97&	7.23&	8.31&	7.73&	8.38\\
HF&	6.23&	6.27&	4.8&	5.52&	4.89&	5.6\\
\hline
$\Delta\,(\%)$  & 6.1 & 5.3  &  9.8   & 2.0   &  8.9   &    \\
\hline
\end{tabular}
  \begin{flushleft}
    $^a$ The last row shows the mean absolute error in percentage.
    $^b$ Experimental results taken from Gr{\"u}ning et
    al.~\cite{gggb02:9591}.
  \end{flushleft} 
\end{table}

For static (isotropic) dipole polarizabilities (see
\ref{table_molecules2}) the situation is different in the sense that
the LDA and PBE perform rather well, which is surprising in view of
the fact that the polarization is largely a non-local and collective
effect. It is noteworthy, however, that the present test set does not
include problematic elongated molecules or chains (see next section),
for which going beyond LDA (and GGA) is
essential~\cite{kkp04:213002,gggb02:9591,mwy03:11001,nf07:191106,kak09:712,rpcsv08:060502,mwdrs09:300,ggb02:6435}.
For the present cases BJBR works remarkably well with a mean error of
only 2\%, whereas RPPBR and LB94 deviate almost 10\% from the
experiments. Nevertheless, no dramatic failures are obtained by using
any of the tested approximations.

\subsection{Hydrogen chains}

\begin{turnpage}
\begin{table}
  \caption{\label{table_chains} Longitudinal polarizabilities of
    hydrogen chains (in a.u.)$^a$}
  \begin{tabular}{l c c c c c c c c c c c c}
\hline
Chain   & xLDA&  xPBE&  BJBR& RPPBR&   OEP$^b$&   LDA&   PBE&  LB94& BJBR-PW& RPPBR-PW& CCSD(T)$^c$& MP4$^c$\\
\hline 
H$_2$   &  13.1&  12.5&  12.4&  11.2&      &  12.4&  12.0&  11.2&      11.8&       10.8&        &       \\
H$_4$   &  39.6&  37.2&  36.3&  33.3&  32.2&  37.7&  36.1&  35.5&      34.9&       32.4&      29&  29.5 \\
H$_6$   &  76,4&  70.7&  68.6&  63.6&  65.6&  72.9&  69.4&  70.5&      65.8&       61.6&    50.9&  51.9 \\
H$_8$   & 120.6& 110.2& 106.0&  99.0&  84.2& 115.2& 108.8& 112.9&     101.6&       95.8&    74.4&  76.2 \\
H$_{10}$& 169.9& 153.2& 146.1& 137.1&      & 162.2& 152.1& 160.5&     140.8&      132.7&        &       \\	
H$_{12}$& 222.4& 199.2& 188.4& 177.2& 138.1& 212.2& 197.8& 211.6&     182.1&      171.1&     124& 127.3 \\
H$_{14}$& 277.0& 246.1& 231.9& 218.0&      & 264.0& 245.2& 264.3&     224.1&      210.6&        & 155   \\
H$_{16}$& 333.0& 294.1& 277.5& 259.5&      & 317.2& 293.4& 318.6&     267.3&      250.5&        &       \\
H$_{18}$& 389.8& 342.5& 323.0& 301.5&      & 371.1& 342.2& 373.2&     309.8&      290.8&        & 205.39\\
H$_{20}$& 447.3& 391.4& 367.2& 343.6&      & 425.4& 391.4& 425.0&     353.9&      331.3&        &       \\		
\hline
$\Delta\,(\%)$& 40.6& 28.9& 24.0& 15.4&    &  56.2&  46.5&  53.8&      36.2&       27.7&        &  \\
\hline
\end{tabular}
  \begin{flushleft}
    $^a$ The last row shows the mean absolute error in percentage,
    calculated against OEP and MP4 (when available) for exchange only
    potentials and combined exchange and correlation potentials,
    respectively.  $^b$ Results from the work of K\"ummel et
    al.~\cite{kkp04:213002} $^c$ The MP4 and CCSD(T) results have been
    taken from the work of Ruzsinszky et al.~\cite{rpcsv08:060502}
    apart from the MP4 result for H$_{18}$ taken from Champagne et
    al.~\cite{cmva95:178}.
  \end{flushleft} 
\end{table}
\end{turnpage}

In \ref{table_chains} we show the polarizabilities calculated for
hydrogen chains from H$_2$ up to H$_{20}$. As the reference results we
use available data from the CCSD(T) (coupled-cluster with single and
double and perturbative triple excitations) and MP4 (fourth-order
M{\o}ller-Plesset perturbation theory)~\cite{rpcsv08:060502}. This
well-studied system has proved to be a remarkable challenge for
DFT~\cite{kkp04:213002,rpcsv08:060502,mwdrs09:300,ggb02:6435,akk08:165106}.
For example, LDA severely overestimates the polarizability, as
demonstrated also by our results in \ref{table_chains}.  The error of
PBE is slightly smaller. The failure of LDA and PBE to capture the
electric response is believed to be due to the inherent
self-interaction
error~\cite{rpcsv08:060502,psb08:121204,kmk08:133004}. We find that
the mean error of LB94 is almost the same as the one of LDA, whereas
for BJBR it is smaller. RPPBR has the best performance of all the
tested potentials when compared to MP4, although it is still quite
large (27.7\%). Possible sources of error in RPPBR (and BJBR) results
are the ultra-non-local effects in long chains, which might be beyond
reach of any semi-local functionals without {\em ad hoc}
modifications, and the using of LDA for the correlation part. This
last point seems to be confirmed by the results obtained without
adding a correlation part to the exchange potentials: when comparing
the polarizabilities obtained from the exchange-only potentials
against exact-exchange OEP results~\cite{kkp04:213002}, all the
average errors are reduced, while the overall trend remains the same.

\section{Summary and outlook}

In summary, we have tested recently constructed
meta-generalized-gradient (meta-GGA) functionals for the exchange
potential, in particular the potential of R\"as\"anen, Pittalis, and
Proetto (RPP) and that of Becke and Johnson (BJ), when complemented by
the Becke-Roussel (BR) approximation to the Slater potential (denoted
in total as RPPBR and BJBR, respectively), and by the correlation in
the LDA level.  These approximations were compared to the van Leeuwen
and Baerends potential (LB94), a GGA that shares some properties with
these new meta-GGAs, as well as to standard LDA and GGA
functionals. As the reference data we used experimental results
whenever available, numerically exact data, and, in the case of
comparing the exchange-only results, the exact-exchange results
obtained from the optimized-effective-potential method.

Overall, the RPPBR potential fared best in the present testsuite
consisting of ionization potentials and electronic affinities of
atoms, ionization potentials and dipole polarizabilities of small
molecules, and longitudinal polarizabilities of hydrogen chains.  LB94
potential performed in an appealing fashion in several instances.  The
BJBR potential gave particularly good results for the calculation of
static polarizabilities of small molecules. Desired future
developments would include the development of correlation potentials
compatible with the RPRBR potential.

In conclusion, the RPPBR potential combines a proper theoretical
foundation with very good results for a series of properties of atoms
and molecules. Moreover, it is very light from the computational point
of view, thus allowing an efficient calculation of large systems.
Therefore, we believe that the RPPBR potential is an important step in
the quest for a simple and all-round semi-local potential for
applications of density-functional theory.

\section*{Acknowledgements}

This work was supported by the Academy of Finland, and the EU's Sixth
Framework Programme through the ETSF e-I3.  SP acknowledges support by
DOE grant DE-FG02-05ER46203. MO thankfully acknowledges financial
support from the Portuguese FCT (contract \#SFRH/BPD/44608/2008). MALM
acknowledges partial funding from the French ANR
(ANR-08-CEXC8-008-01), and from the program PIR Mat\'eriaux -- MaProSu
of CNRS. Part of the calculations were performed at the LCA of the
University of Coimbra and at GENCI (project x2010096017).

\bibliography{rpp}

\begin{thebibliography}{50}
\expandafter\ifx\csname natexlab\endcsname\relax\def\natexlab#1{#1}\fi
\expandafter\ifx\csname bibnamefont\endcsname\relax
  \def\bibnamefont#1{#1}\fi
\expandafter\ifx\csname bibfnamefont\endcsname\relax
  \def\bibfnamefont#1{#1}\fi
\expandafter\ifx\csname citenamefont\endcsname\relax
  \def\citenamefont#1{#1}\fi
\expandafter\ifx\csname url\endcsname\relax
  \def\url#1{\texttt{#1}}\fi
\expandafter\ifx\csname urlprefix\endcsname\relax\def\urlprefix{URL }\fi
\providecommand{\bibinfo}[2]{#2}
\providecommand{\eprint}[2][]{\url{#2}}

\bibitem[{\citenamefont{Dreizler and Gross}(1990)}]{dg90}
\bibinfo{author}{\bibfnamefont{R.~M.} \bibnamefont{Dreizler}} \bibnamefont{and}
  \bibinfo{author}{\bibfnamefont{E.~K.~U.} \bibnamefont{Gross}},
  \emph{\bibinfo{title}{Density Functional Theory}}
  (\bibinfo{publisher}{Springer}, \bibinfo{address}{Berlin},
  \bibinfo{year}{1990}).

\bibitem[{\citenamefont{von Barth}(2004)}]{barth04:9}
\bibinfo{author}{\bibfnamefont{U.}~\bibnamefont{von Barth}},
  \bibinfo{journal}{Phys. Scr.} \textbf{\bibinfo{volume}{109}},
  \bibinfo{pages}{9} (\bibinfo{year}{2004}).

\bibitem[{\citenamefont{Perdew and Kurth}(2003)}]{pk03}
\bibinfo{author}{\bibfnamefont{J.~P.} \bibnamefont{Perdew}} \bibnamefont{and}
  \bibinfo{author}{\bibfnamefont{S.}~\bibnamefont{Kurth}},
  \emph{\bibinfo{title}{A Primer in Density Functional Theory}}
  (\bibinfo{publisher}{Springer}, \bibinfo{address}{Berlin},
  \bibinfo{year}{2003}), vol. \bibinfo{volume}{620} of
  \emph{\bibinfo{series}{Lecture Notes in Physics}}, pp.
  \bibinfo{pages}{1--55}.

\bibitem[{\citenamefont{Heyd et~al.}(2005)\citenamefont{Heyd, Peralta,
  Scuseria, and Martin}}]{hpsm05:174101}
\bibinfo{author}{\bibfnamefont{J.}~\bibnamefont{Heyd}},
  \bibinfo{author}{\bibfnamefont{J.~E.} \bibnamefont{Peralta}},
  \bibinfo{author}{\bibfnamefont{G.~E.} \bibnamefont{Scuseria}},
  \bibnamefont{and} \bibinfo{author}{\bibfnamefont{R.~L.}
  \bibnamefont{Martin}}, \bibinfo{journal}{J. Chem. Phys.}
  \textbf{\bibinfo{volume}{123}}, \bibinfo{pages}{174101}
  (\bibinfo{year}{2005}).

\bibitem[{\citenamefont{Champagne et~al.}(1998)\citenamefont{Champagne,
  Perp{\`e}te, van Gisbergen, Baerends, Snijders, Soubra-Ghaoui, Robins, and
  Kirtman}}]{cpgbssrk98:10489}
\bibinfo{author}{\bibfnamefont{B.}~\bibnamefont{Champagne}},
  \bibinfo{author}{\bibfnamefont{E.~A.} \bibnamefont{Perp{\`e}te}},
  \bibinfo{author}{\bibfnamefont{S.~J.~A.} \bibnamefont{van Gisbergen}},
  \bibinfo{author}{\bibfnamefont{E.-J.} \bibnamefont{Baerends}},
  \bibinfo{author}{\bibfnamefont{J.~G.} \bibnamefont{Snijders}},
  \bibinfo{author}{\bibfnamefont{C.}~\bibnamefont{Soubra-Ghaoui}},
  \bibinfo{author}{\bibfnamefont{K.~A.} \bibnamefont{Robins}},
  \bibnamefont{and} \bibinfo{author}{\bibfnamefont{B.}~\bibnamefont{Kirtman}},
  \bibinfo{journal}{J. Chem. Phys.} \textbf{\bibinfo{volume}{109}},
  \bibinfo{pages}{10489} (\bibinfo{year}{1998}).

\bibitem[{\citenamefont{K{\"u}mmel et~al.}(2004)\citenamefont{K{\"u}mmel,
  Kronik, and Perdew}}]{kkp04:213002}
\bibinfo{author}{\bibfnamefont{S.}~\bibnamefont{K{\"u}mmel}},
  \bibinfo{author}{\bibfnamefont{L.}~\bibnamefont{Kronik}}, \bibnamefont{and}
  \bibinfo{author}{\bibfnamefont{J.~P.} \bibnamefont{Perdew}},
  \bibinfo{journal}{Phys. Rev. Lett.} \textbf{\bibinfo{volume}{93}},
  \bibinfo{pages}{213002} (\bibinfo{year}{2004}).

\bibitem[{\citenamefont{Gr{\"u}ning
  et~al.}(2002{\natexlab{a}})\citenamefont{Gr{\"u}ning, Gritsenko, van
  Gisbergen, and Baerends}}]{gggb02:9591}
\bibinfo{author}{\bibfnamefont{M.}~\bibnamefont{Gr{\"u}ning}},
  \bibinfo{author}{\bibfnamefont{O.~V.} \bibnamefont{Gritsenko}},
  \bibinfo{author}{\bibfnamefont{S.~J.~A.} \bibnamefont{van Gisbergen}},
  \bibnamefont{and} \bibinfo{author}{\bibfnamefont{E.~J.}
  \bibnamefont{Baerends}}, \bibinfo{journal}{J. Chem. Phys.}
  \textbf{\bibinfo{volume}{116}}, \bibinfo{pages}{9591}
  (\bibinfo{year}{2002}{\natexlab{a}}).

\bibitem[{\citenamefont{Mori-S{\'a}nchez
  et~al.}(2003)\citenamefont{Mori-S{\'a}nchez, Wu, and Yang}}]{mwy03:11001}
\bibinfo{author}{\bibfnamefont{P.}~\bibnamefont{Mori-S{\'a}nchez}},
  \bibinfo{author}{\bibfnamefont{Q.}~\bibnamefont{Wu}}, \bibnamefont{and}
  \bibinfo{author}{\bibfnamefont{W.}~\bibnamefont{Yang}}, \bibinfo{journal}{J.
  Chem. Phys.} \textbf{\bibinfo{volume}{119}}, \bibinfo{pages}{11001}
  (\bibinfo{year}{2003}).

\bibitem[{\citenamefont{Maitra and van Faassen}(2007)}]{nf07:191106}
\bibinfo{author}{\bibfnamefont{N.~T.} \bibnamefont{Maitra}} \bibnamefont{and}
  \bibinfo{author}{\bibfnamefont{M.}~\bibnamefont{van Faassen}},
  \bibinfo{journal}{J. Chem. Phys.} \textbf{\bibinfo{volume}{126}},
  \bibinfo{pages}{191106} (\bibinfo{year}{2007}).

\bibitem[{\citenamefont{Karolewski et~al.}(2009)\citenamefont{Karolewski,
  Armiento, and K{\"u}mmel}}]{kak09:712}
\bibinfo{author}{\bibfnamefont{A.}~\bibnamefont{Karolewski}},
  \bibinfo{author}{\bibfnamefont{R.}~\bibnamefont{Armiento}}, \bibnamefont{and}
  \bibinfo{author}{\bibfnamefont{S.}~\bibnamefont{K{\"u}mmel}},
  \bibinfo{journal}{J. Chem. Theory Comput.} \textbf{\bibinfo{volume}{5}},
  \bibinfo{pages}{712} (\bibinfo{year}{2009}).

\bibitem[{\citenamefont{Ruzsinszky et~al.}(2008)\citenamefont{Ruzsinszky,
  Perdew, Csonka, Scuseria, and Vydrov;}}]{rpcsv08:060502}
\bibinfo{author}{\bibfnamefont{A.}~\bibnamefont{Ruzsinszky}},
  \bibinfo{author}{\bibfnamefont{J.~P.} \bibnamefont{Perdew}},
  \bibinfo{author}{\bibfnamefont{G.~I.} \bibnamefont{Csonka}},
  \bibinfo{author}{\bibfnamefont{G.~E.} \bibnamefont{Scuseria}},
  \bibnamefont{and} \bibinfo{author}{\bibfnamefont{O.~A.}
  \bibnamefont{Vydrov;}}, \bibinfo{journal}{Phys. Rev. A}
  \textbf{\bibinfo{volume}{77}}, \bibinfo{pages}{060502(R)}
  (\bibinfo{year}{2008}).

\bibitem[{\citenamefont{Messud et~al.}(2009)\citenamefont{Messud, Wang, Dinh,
  Reinhard, and Suraud}}]{mwdrs09:300}
\bibinfo{author}{\bibfnamefont{J.}~\bibnamefont{Messud}},
  \bibinfo{author}{\bibfnamefont{Z.}~\bibnamefont{Wang}},
  \bibinfo{author}{\bibfnamefont{P.~M.} \bibnamefont{Dinh}},
  \bibinfo{author}{\bibfnamefont{P.-G.} \bibnamefont{Reinhard}},
  \bibnamefont{and} \bibinfo{author}{\bibfnamefont{E.}~\bibnamefont{Suraud}},
  \bibinfo{journal}{Chem. Phys. Lett.} \textbf{\bibinfo{volume}{479}},
  \bibinfo{pages}{300} (\bibinfo{year}{2009}).

\bibitem[{\citenamefont{Gr{\"u}ning
  et~al.}(2002{\natexlab{b}})\citenamefont{Gr{\"u}ning, Gritsenko, and
  Baerends}}]{ggb02:6435}
\bibinfo{author}{\bibfnamefont{M.}~\bibnamefont{Gr{\"u}ning}},
  \bibinfo{author}{\bibfnamefont{O.~V.} \bibnamefont{Gritsenko}},
  \bibnamefont{and} \bibinfo{author}{\bibfnamefont{E.~J.}
  \bibnamefont{Baerends}}, \bibinfo{journal}{J. Chem. Phys.}
  \textbf{\bibinfo{volume}{116}}, \bibinfo{pages}{6435}
  (\bibinfo{year}{2002}{\natexlab{b}}).

\bibitem[{\citenamefont{Sharp and Horton}(1953)}]{sh53:317}
\bibinfo{author}{\bibfnamefont{R.~T.} \bibnamefont{Sharp}} \bibnamefont{and}
  \bibinfo{author}{\bibfnamefont{G.~K.} \bibnamefont{Horton}},
  \bibinfo{journal}{Phys. Rev.} \textbf{\bibinfo{volume}{90}},
  \bibinfo{pages}{317} (\bibinfo{year}{1953}).

\bibitem[{\citenamefont{Talman and Shadwick}(1976)}]{ts76:36}
\bibinfo{author}{\bibfnamefont{J.~D.} \bibnamefont{Talman}} \bibnamefont{and}
  \bibinfo{author}{\bibfnamefont{W.~F.} \bibnamefont{Shadwick}},
  \bibinfo{journal}{Phys. Rev. A} \textbf{\bibinfo{volume}{14}},
  \bibinfo{pages}{36} (\bibinfo{year}{1976}).

\bibitem[{\citenamefont{K{\"u}mmel and Kronik}(2008)}]{kk08:3}
\bibinfo{author}{\bibfnamefont{S.}~\bibnamefont{K{\"u}mmel}} \bibnamefont{and}
  \bibinfo{author}{\bibfnamefont{L.}~\bibnamefont{Kronik}},
  \bibinfo{journal}{Rev. Mod. Phys.} \textbf{\bibinfo{volume}{80}},
  \bibinfo{pages}{3} (\bibinfo{year}{2008}).

\bibitem[{\citenamefont{Krieger et~al.}(1992)\citenamefont{Krieger, Li, and
  Iafrate}}]{kli92:101}
\bibinfo{author}{\bibfnamefont{J.~B.} \bibnamefont{Krieger}},
  \bibinfo{author}{\bibfnamefont{Y.}~\bibnamefont{Li}}, \bibnamefont{and}
  \bibinfo{author}{\bibfnamefont{G.~J.} \bibnamefont{Iafrate}},
  \bibinfo{journal}{Phys. Rev. A} \textbf{\bibinfo{volume}{45}},
  \bibinfo{pages}{101} (\bibinfo{year}{1992}).

\bibitem[{\citenamefont{Perdew et~al.}(1999)\citenamefont{Perdew, Kurth, Zupan,
  and Blaha}}]{pkzb99:2544}
\bibinfo{author}{\bibfnamefont{J.~P.} \bibnamefont{Perdew}},
  \bibinfo{author}{\bibfnamefont{S.}~\bibnamefont{Kurth}},
  \bibinfo{author}{\bibfnamefont{A.}~\bibnamefont{Zupan}}, \bibnamefont{and}
  \bibinfo{author}{\bibfnamefont{P.}~\bibnamefont{Blaha}},
  \bibinfo{journal}{Phys. Rev. Lett.} \textbf{\bibinfo{volume}{82}},
  \bibinfo{pages}{2544} (\bibinfo{year}{1999}).

\bibitem[{\citenamefont{Tao et~al.}(2003)\citenamefont{Tao, Perdew, Staroverov,
  and Scuseria}}]{tpss03:146401}
\bibinfo{author}{\bibfnamefont{J.}~\bibnamefont{Tao}},
  \bibinfo{author}{\bibfnamefont{J.~P.} \bibnamefont{Perdew}},
  \bibinfo{author}{\bibfnamefont{V.~N.} \bibnamefont{Staroverov}},
  \bibnamefont{and} \bibinfo{author}{\bibfnamefont{G.~E.}
  \bibnamefont{Scuseria}}, \bibinfo{journal}{Phys. Rev. Lett.}
  \textbf{\bibinfo{volume}{91}}, \bibinfo{pages}{146401}
  (\bibinfo{year}{2003}).

\bibitem[{\citenamefont{R{\"a}s{\"a}nen
  et~al.}(2010{\natexlab{a}})\citenamefont{R{\"a}s{\"a}nen, Pittalis, and
  Proetto}}]{rpp10:044112}
\bibinfo{author}{\bibfnamefont{E.}~\bibnamefont{R{\"a}s{\"a}nen}},
  \bibinfo{author}{\bibfnamefont{S.}~\bibnamefont{Pittalis}}, \bibnamefont{and}
  \bibinfo{author}{\bibfnamefont{C.~R.} \bibnamefont{Proetto}},
  \bibinfo{journal}{J. Chem. Phys.} \textbf{\bibinfo{volume}{132}},
  \bibinfo{pages}{044112} (\bibinfo{year}{2010}{\natexlab{a}}).

\bibitem[{\citenamefont{Becke and Johnson}(2006)}]{bj06:221101}
\bibinfo{author}{\bibfnamefont{A.~D.} \bibnamefont{Becke}} \bibnamefont{and}
  \bibinfo{author}{\bibfnamefont{E.~R.} \bibnamefont{Johnson}},
  \bibinfo{journal}{J. Chem. Phys.} \textbf{\bibinfo{volume}{124}},
  \bibinfo{pages}{221101} (\bibinfo{year}{2006}).

\bibitem[{\citenamefont{Armiento et~al.}(2008)\citenamefont{Armiento,
  K{\"u}mmel, and K{\"o}rzd{\"o}rfer}}]{akk08:165106}
\bibinfo{author}{\bibfnamefont{R.}~\bibnamefont{Armiento}},
  \bibinfo{author}{\bibfnamefont{S.}~\bibnamefont{K{\"u}mmel}},
  \bibnamefont{and}
  \bibinfo{author}{\bibfnamefont{T.}~\bibnamefont{K{\"o}rzd{\"o}rfer}},
  \bibinfo{journal}{Phys. Rev. B} \textbf{\bibinfo{volume}{77}},
  \bibinfo{pages}{165106} (\bibinfo{year}{2008}).

\bibitem[{\citenamefont{Tran and Blaha}(2009)}]{tb09:226401}
\bibinfo{author}{\bibfnamefont{F.}~\bibnamefont{Tran}} \bibnamefont{and}
  \bibinfo{author}{\bibfnamefont{P.}~\bibnamefont{Blaha}},
  \bibinfo{journal}{Phys. Rev. Lett.} \textbf{\bibinfo{volume}{102}},
  \bibinfo{pages}{226401} (\bibinfo{year}{2009}).

\bibitem[{\citenamefont{Becke and Roussel}(1989)}]{br89:3761}
\bibinfo{author}{\bibfnamefont{A.~D.} \bibnamefont{Becke}} \bibnamefont{and}
  \bibinfo{author}{\bibfnamefont{M.~R.} \bibnamefont{Roussel}},
  \bibinfo{journal}{Phys. Rev. A} \textbf{\bibinfo{volume}{39}},
  \bibinfo{pages}{3761} (\bibinfo{year}{1989}).

\bibitem[{\citenamefont{van Leeuwen and Baerends}(1994)}]{lb94:2421}
\bibinfo{author}{\bibfnamefont{R.}~\bibnamefont{van Leeuwen}} \bibnamefont{and}
  \bibinfo{author}{\bibfnamefont{E.~J.} \bibnamefont{Baerends}},
  \bibinfo{journal}{Phys. Rev. A} \textbf{\bibinfo{volume}{49}},
  \bibinfo{pages}{2421} (\bibinfo{year}{1994}).

\bibitem[{\citenamefont{Dobson}(1993)}]{dobson93:8870}
\bibinfo{author}{\bibfnamefont{J.~F.} \bibnamefont{Dobson}},
  \bibinfo{journal}{J. Chem. Phys.} \textbf{\bibinfo{volume}{98}},
  \bibinfo{pages}{8870} (\bibinfo{year}{1993}).

\bibitem[{\citenamefont{Becke}(1988)}]{becke88:1053}
\bibinfo{author}{\bibfnamefont{A.~D.} \bibnamefont{Becke}},
  \bibinfo{journal}{J. Chem. Phys.} \textbf{\bibinfo{volume}{88}},
  \bibinfo{pages}{1053} (\bibinfo{year}{1988}).

\bibitem[{\citenamefont{Becke}(1996)}]{becke96:995}
\bibinfo{author}{\bibfnamefont{A.~D.} \bibnamefont{Becke}},
  \bibinfo{journal}{Can. J. Chem.} \textbf{\bibinfo{volume}{74}},
  \bibinfo{pages}{995} (\bibinfo{year}{1996}).

\bibitem[{\citenamefont{Pittalis et~al.}(2007)\citenamefont{Pittalis,
  R{\"a}s{\"a}nen, Helbig, and Gross}}]{prhg07:235314}
\bibinfo{author}{\bibfnamefont{S.}~\bibnamefont{Pittalis}},
  \bibinfo{author}{\bibfnamefont{E.}~\bibnamefont{R{\"a}s{\"a}nen}},
  \bibinfo{author}{\bibfnamefont{N.}~\bibnamefont{Helbig}}, \bibnamefont{and}
  \bibinfo{author}{\bibfnamefont{E.~K.~U.} \bibnamefont{Gross}},
  \bibinfo{journal}{Phys. Rev. B} \textbf{\bibinfo{volume}{76}},
  \bibinfo{pages}{235314} (\bibinfo{year}{2007}).

\bibitem[{\citenamefont{Pittalis et~al.}(2009)\citenamefont{Pittalis,
  R{\"a}s{\"a}nen, Proetto, and Gross}}]{prpg09:085316}
\bibinfo{author}{\bibfnamefont{S.}~\bibnamefont{Pittalis}},
  \bibinfo{author}{\bibfnamefont{E.}~\bibnamefont{R{\"a}s{\"a}nen}},
  \bibinfo{author}{\bibfnamefont{C.~R.} \bibnamefont{Proetto}},
  \bibnamefont{and} \bibinfo{author}{\bibfnamefont{E.~K.~U.}
  \bibnamefont{Gross}}, \bibinfo{journal}{Phys. Rev. B}
  \textbf{\bibinfo{volume}{79}}, \bibinfo{pages}{085316}
  (\bibinfo{year}{2009}).

\bibitem[{\citenamefont{R{\"a}s{\"a}nen
  et~al.}(2010{\natexlab{b}})\citenamefont{R{\"a}s{\"a}nen, Pittalis, and
  Proetto}}]{rpp10:195103}
\bibinfo{author}{\bibfnamefont{E.}~\bibnamefont{R{\"a}s{\"a}nen}},
  \bibinfo{author}{\bibfnamefont{S.}~\bibnamefont{Pittalis}}, \bibnamefont{and}
  \bibinfo{author}{\bibfnamefont{C.~R.} \bibnamefont{Proetto}},
  \bibinfo{journal}{Phys. Rev. B} \textbf{\bibinfo{volume}{81}},
  \bibinfo{pages}{195103} (\bibinfo{year}{2010}{\natexlab{b}}).

\bibitem[{\citenamefont{Becke and Edgecombe}(1990)}]{bh90:5397}
\bibinfo{author}{\bibfnamefont{A.~D.} \bibnamefont{Becke}} \bibnamefont{and}
  \bibinfo{author}{\bibfnamefont{K.~E.} \bibnamefont{Edgecombe}},
  \bibinfo{journal}{J. Chem. Phys.} \textbf{\bibinfo{volume}{92}},
  \bibinfo{pages}{5397} (\bibinfo{year}{1990}).

\bibitem[{\citenamefont{Burnus et~al.}(2005)\citenamefont{Burnus, Marques, and
  Gross}}]{bmg05:010501}
\bibinfo{author}{\bibfnamefont{T.}~\bibnamefont{Burnus}},
  \bibinfo{author}{\bibfnamefont{M.~A.~L.} \bibnamefont{Marques}},
  \bibnamefont{and} \bibinfo{author}{\bibfnamefont{E.~K.~U.}
  \bibnamefont{Gross}}, \bibinfo{journal}{Phys. Rev. A}
  \textbf{\bibinfo{volume}{71}}, \bibinfo{pages}{010501(R)}
  (\bibinfo{year}{2005}).

\bibitem[{\citenamefont{R{\"a}s{\"a}nen
  et~al.}(2008)\citenamefont{R{\"a}s{\"a}nen, Castro, and
  Gross}}]{rcg08:115108}
\bibinfo{author}{\bibfnamefont{E.}~\bibnamefont{R{\"a}s{\"a}nen}},
  \bibinfo{author}{\bibfnamefont{A.}~\bibnamefont{Castro}}, \bibnamefont{and}
  \bibinfo{author}{\bibfnamefont{E.~K.~U.} \bibnamefont{Gross}},
  \bibinfo{journal}{Phys. Rev. B} \textbf{\bibinfo{volume}{77}},
  \bibinfo{pages}{115108} (\bibinfo{year}{2008}).

\bibitem[{\citenamefont{Sala and G{\"o}rling}(2002)}]{sg02:033003}
\bibinfo{author}{\bibfnamefont{F.~D.} \bibnamefont{Sala}} \bibnamefont{and}
  \bibinfo{author}{\bibfnamefont{A.}~\bibnamefont{G{\"o}rling}},
  \bibinfo{journal}{Phys. Rev. Lett.} \textbf{\bibinfo{volume}{89}},
  \bibinfo{pages}{033003} (\bibinfo{year}{2002}).

\bibitem[{\citenamefont{K{\"u}mmel and Perdew}(2003)}]{kp03:035103}
\bibinfo{author}{\bibfnamefont{S.}~\bibnamefont{K{\"u}mmel}} \bibnamefont{and}
  \bibinfo{author}{\bibfnamefont{J.~P.} \bibnamefont{Perdew}},
  \bibinfo{journal}{Phys. Rev. B} \textbf{\bibinfo{volume}{68}},
  \bibinfo{pages}{035103} (\bibinfo{year}{2003}).

\bibitem[{\citenamefont{Pittalis et~al.}(2010)\citenamefont{Pittalis,
  R{\"a}s{\"a}nen, and Proetto}}]{prp10:115108}
\bibinfo{author}{\bibfnamefont{S.}~\bibnamefont{Pittalis}},
  \bibinfo{author}{\bibfnamefont{E.}~\bibnamefont{R{\"a}s{\"a}nen}},
  \bibnamefont{and} \bibinfo{author}{\bibfnamefont{C.~R.}
  \bibnamefont{Proetto}}, \bibinfo{journal}{Phys. Rev. B}
  \textbf{\bibinfo{volume}{81}}, \bibinfo{pages}{115108}
  (\bibinfo{year}{2010}).

\bibitem[{\citenamefont{Perdew and Wang}(1992)}]{pw92:13244}
\bibinfo{author}{\bibfnamefont{J.~P.} \bibnamefont{Perdew}} \bibnamefont{and}
  \bibinfo{author}{\bibfnamefont{Y.}~\bibnamefont{Wang}},
  \bibinfo{journal}{Phys. Rev. B} \textbf{\bibinfo{volume}{45}},
  \bibinfo{pages}{13244} (\bibinfo{year}{1992}).

\bibitem[{\citenamefont{Perdew et~al.}(1996)\citenamefont{Perdew, Burke, and
  Ernzerhof}}]{pbe96:3865}
\bibinfo{author}{\bibfnamefont{J.~P.} \bibnamefont{Perdew}},
  \bibinfo{author}{\bibfnamefont{K.}~\bibnamefont{Burke}}, \bibnamefont{and}
  \bibinfo{author}{\bibfnamefont{M.}~\bibnamefont{Ernzerhof}},
  \bibinfo{journal}{Phys. Rev. Lett.} \textbf{\bibinfo{volume}{77}},
  \bibinfo{pages}{3865} (\bibinfo{year}{1996}).

\bibitem[{\citenamefont{Gaiduk and Staroverov}(2009)}]{gs09:044107}
\bibinfo{author}{\bibfnamefont{A.~P.} \bibnamefont{Gaiduk}} \bibnamefont{and}
  \bibinfo{author}{\bibfnamefont{V.~N.} \bibnamefont{Staroverov}},
  \bibinfo{journal}{J. Chem. Phys.} \textbf{\bibinfo{volume}{131}},
  \bibinfo{pages}{044107} (\bibinfo{year}{2009}).

\bibitem[{\citenamefont{Oliveira and Nogueira}(2008)}]{on08:524}
\bibinfo{author}{\bibfnamefont{M.~J.~T.} \bibnamefont{Oliveira}}
  \bibnamefont{and} \bibinfo{author}{\bibfnamefont{F.}~\bibnamefont{Nogueira}},
  \bibinfo{journal}{Comp. Phys. Comm.} \textbf{\bibinfo{volume}{178}},
  \bibinfo{pages}{524} (\bibinfo{year}{2008}).

\bibitem[{\citenamefont{Marques et~al.}(2003)\citenamefont{Marques, Castro,
  Bertsch, and Rubio}}]{mcbr03:60}
\bibinfo{author}{\bibfnamefont{M.~A.~L.} \bibnamefont{Marques}},
  \bibinfo{author}{\bibfnamefont{A.}~\bibnamefont{Castro}},
  \bibinfo{author}{\bibfnamefont{G.~F.} \bibnamefont{Bertsch}},
  \bibnamefont{and} \bibinfo{author}{\bibfnamefont{A.}~\bibnamefont{Rubio}},
  \bibinfo{journal}{Comp. Phys. Comm.} \textbf{\bibinfo{volume}{151}},
  \bibinfo{pages}{60} (\bibinfo{year}{2003}).

\bibitem[{\citenamefont{Castro et~al.}(2006)\citenamefont{Castro, Appel,
  Oliveira, Rozzi, Andrade, Lorenzen, Marques, Gross, and
  Rubio}}]{caoralmgr06:2465}
\bibinfo{author}{\bibfnamefont{A.}~\bibnamefont{Castro}},
  \bibinfo{author}{\bibfnamefont{H.}~\bibnamefont{Appel}},
  \bibinfo{author}{\bibfnamefont{M.}~\bibnamefont{Oliveira}},
  \bibinfo{author}{\bibfnamefont{C.~A.} \bibnamefont{Rozzi}},
  \bibinfo{author}{\bibfnamefont{X.}~\bibnamefont{Andrade}},
  \bibinfo{author}{\bibfnamefont{F.}~\bibnamefont{Lorenzen}},
  \bibinfo{author}{\bibfnamefont{M.~A.~L.} \bibnamefont{Marques}},
  \bibinfo{author}{\bibfnamefont{E.~K.~U.} \bibnamefont{Gross}},
  \bibnamefont{and} \bibinfo{author}{\bibfnamefont{A.}~\bibnamefont{Rubio}},
  \bibinfo{journal}{Phys. Stat. Sol. B} \textbf{\bibinfo{volume}{243}},
  \bibinfo{pages}{2465} (\bibinfo{year}{2006}).

\bibitem[{\citenamefont{Colle and Salvetti}(1975)}]{cs75:329}
\bibinfo{author}{\bibfnamefont{R.}~\bibnamefont{Colle}} \bibnamefont{and}
  \bibinfo{author}{\bibfnamefont{O.}~\bibnamefont{Salvetti}},
  \bibinfo{journal}{Theor. Chim. Acta} \textbf{\bibinfo{volume}{37}},
  \bibinfo{pages}{329} (\bibinfo{year}{1975}).

\bibitem[{\citenamefont{Grabo and Gross}(1995)}]{gg95:141}
\bibinfo{author}{\bibfnamefont{T.}~\bibnamefont{Grabo}} \bibnamefont{and}
  \bibinfo{author}{\bibfnamefont{E.~K.~U.} \bibnamefont{Gross}},
  \bibinfo{journal}{Chem. Phys. Lett.} \textbf{\bibinfo{volume}{240}},
  \bibinfo{pages}{141} (\bibinfo{year}{1995}).

\bibitem[{\citenamefont{Engel and Vosko}(1993)}]{ev93:2800}
\bibinfo{author}{\bibfnamefont{E.}~\bibnamefont{Engel}} \bibnamefont{and}
  \bibinfo{author}{\bibfnamefont{S.~H.} \bibnamefont{Vosko}},
  \bibinfo{journal}{Phys. Rev. A} \textbf{\bibinfo{volume}{47}},
  \bibinfo{pages}{2800} (\bibinfo{year}{1993}).

\bibitem[{\citenamefont{Radzig and Smirnov}(1985)}]{rs85}
\bibinfo{author}{\bibfnamefont{A.~A.} \bibnamefont{Radzig}} \bibnamefont{and}
  \bibinfo{author}{\bibfnamefont{B.~M.} \bibnamefont{Smirnov}},
  \emph{\bibinfo{title}{Reference Data on Atoms and Molecules}}
  (\bibinfo{publisher}{Springer Verlag}, \bibinfo{address}{Berlin},
  \bibinfo{year}{1985}).

\bibitem[{\citenamefont{Champagne et~al.}(1995)\citenamefont{Champagne, Mosley,
  Vra{\u{c}}ko, and Andr{\'e}}}]{cmva95:178}
\bibinfo{author}{\bibfnamefont{B.}~\bibnamefont{Champagne}},
  \bibinfo{author}{\bibfnamefont{D.~H.} \bibnamefont{Mosley}},
  \bibinfo{author}{\bibfnamefont{M.}~\bibnamefont{Vra{\u{c}}ko}},
  \bibnamefont{and} \bibinfo{author}{\bibfnamefont{J.-M.}
  \bibnamefont{Andr{\'e}}}, \bibinfo{journal}{Phys. Rev. A}
  \textbf{\bibinfo{volume}{52}}, \bibinfo{pages}{178} (\bibinfo{year}{1995}).

\bibitem[{\citenamefont{Pemmaraju et~al.}(2008)\citenamefont{Pemmaraju,
  Sanvito, and Burke}}]{psb08:121204}
\bibinfo{author}{\bibfnamefont{C.~D.} \bibnamefont{Pemmaraju}},
  \bibinfo{author}{\bibfnamefont{S.}~\bibnamefont{Sanvito}}, \bibnamefont{and}
  \bibinfo{author}{\bibfnamefont{K.}~\bibnamefont{Burke}},
  \bibinfo{journal}{Phys. Rev. B} \textbf{\bibinfo{volume}{77}},
  \bibinfo{pages}{121204(R)} (\bibinfo{year}{2008}).

\bibitem[{\citenamefont{K{\"o}rzd{\"o}rfer
  et~al.}(2008)\citenamefont{K{\"o}rzd{\"o}rfer, Mundt, and
  K{\"u}mmel}}]{kmk08:133004}
\bibinfo{author}{\bibfnamefont{T.}~\bibnamefont{K{\"o}rzd{\"o}rfer}},
  \bibinfo{author}{\bibfnamefont{M.}~\bibnamefont{Mundt}}, \bibnamefont{and}
  \bibinfo{author}{\bibfnamefont{S.}~\bibnamefont{K{\"u}mmel}},
  \bibinfo{journal}{Phys. Rev. Lett.} \textbf{\bibinfo{volume}{100}},
  \bibinfo{pages}{133004} (\bibinfo{year}{2008}).

\end{thebibliography}

\end{document}